\documentclass[notoc,12pt]{JHEP3}

\usepackage{amsmath,amssymb,euscript,array,cite}

\def\N{{\cal N}}

\def\Tr{{\rm Tr}\,}
\def\sst{\scriptscriptstyle}

\def\SU{\text{SU}}

\def\U{\text{U}}

\def\SL{\text{SL}}
\newcommand{\BN}{\boldsymbol{N}}

\def\Dbarslash{\,\,{\raise.15ex\hbox{/}\mkern-12mu {\bar\D}}}
\def\Dslash{\,\,{\raise.15ex\hbox{/}\mkern-12mu \D}}
\def\delslash{\,\,{\raise.15ex\hbox{/}\mkern-9mu \partial}}
\def\delbarslash{\,\,{\raise.15ex\hbox{/}\mkern-9mu {\bar\partial}}}

\newcommand{\EQ}[1]{\begin{equation} #1 \end{equation}}
\newcommand{\AL}[1]{\begin{subequations}\begin{align} #1
\end{align}\end{subequations}}
\newcommand{\SP}[1]{\begin{equation}\begin{split} #1 \end{split}\end{equation}}

\title{New Results from Glueball Superpotentials and 
Matrix Models: the Leigh-Strassler Deformation}
\author{Timothy J.~Hollowood\\
Department of Physics, University of Wales Swansea,
Swansea, SA2 8PP, UK\\
E-mail:  {\tt t.hollowood@swan.ac.uk}}
\preprint{SWAT-361}
\abstract{Using the result of a matrix model computation of the
exact glueball superpotential, we investigate the 
relevant mass perturbations of the Leigh-Strassler 
marginal ``$q$'' deformation of $\N=4$ supersymmetric gauge theory. 
We recall a conjecture for the elliptic superpotential that
describes the theory compactified on a circle and identify this
superpotential as one of the Hamiltonians of the elliptic 
Ruijsenaars-Schneider
integrable system. In the limit that the Leigh-Strassler deformation
is turned off, the integrable system reduces to the elliptic Calogero-Moser
system which describes the ${\cal N}=1^*$ theory. 
Based on these results, we identify the Coulomb branch of the
partially mass-deformed Leigh-Strassler theory as the spectral curve of
the Ruijsenaars-Schneider system. We also show how the Leigh-Strassler
deformation may be obtained by suitably modifying Witten's M theory
brane construction of $\N=2$ theories.}

\begin{document}

\section{Introduction}

The motivation of the present paper is to derive some new results in
supersymmetric gauge theories from
the remarkable developments which relate glueball superpotentials of
$\N=1$ theories to matrix models \cite{DV1,DV2,DV3}.

Leigh and Strassler \cite{LS} discovered that $\N=4$ gauge theory has 
two complex marginal deformations. One of them involves replacing the 
tree-level superpotential, which involves the commutator term, by the
``$q$ deformation'':
\EQ{
W_{\rm cl}=
i\,\Tr\,\Phi[\Phi^+,\Phi^-]\longrightarrow
i\lambda\,\Tr\,\Phi[\Phi^+,\Phi^-]_\beta\ ,
}
where we have defined the $q$-commutator
\EQ{[\Phi^+,\Phi^-]_\beta\equiv\Phi^+\Phi^-e^{i\beta/2}-
\Phi^-\Phi^+e^{-i\beta/2}\ .
}
In the above, $\Phi$ and $\Phi^\pm$ are $\SU(N)$ adjoint-valued chiral
fields (unlike \cite{cft} we will only consider the $\SU(N)$ theory
and so we will drop the hats from $\SU(N)$-valued fields).
The resulting theory is known to be finite on some 2-complex
dimensional surface in the space spanned by the two new couplings
$\lambda$ and $\beta$ along with the complex gauge coupling
$\tau$. Away from the $\N=4$ line, $\lambda=1$, $\beta=0$ with $\tau$
arbitrary, the theory only has $\N=1$ supersymmetry. 

In \cite{cft}, generalizing the analysis of the $\N=1^*$ theory in
\cite{DV3,mm1,mm2}, we analysed certain relevant deformations of this space
of theories by using the matrix formalism developed by Dijkgraaf and
Vafa for calculating the exact glueball superpotentials of $\N=1$
theories \cite{DV1,DV2,DV3}. 
The perturbed theory is described by the tree-level superpotential 
\EQ{W_{\rm cl}=
\Tr\big(i\lambda\Phi[\Phi^+,\Phi^-]_\beta+m\Phi^+\Phi^-
+\mu\Phi^2 \big)\ .\label{tree}
}
More generally one can replace the last term by an arbitrary function
$W(\Phi)$. Rather remarkably in the confining vacuum the one-cut
saddle-point solution of the matrix model is solvable thanks to
\cite{kostov}. In \cite{cft}, we showed how the matrix model could
also be solved around certain multi-cut solutions that describe all
the massive vacua of the theory. Explicit formulae will be given
later; however, on the basis of these explicit results we were able to
conjecture an exact form for the ``elliptic'' superpotential of the
theory. This is a superpotential induced by the final term in
\eqref{tree} when the theory
is compactified to three dimensions. It is a rather
useful quantity because, unlike the matrix model glueball
superpotentials, it captures all the vacua in one go and the values of
the condensates that are extracted from it are valid in the
four-dimensional decompactification limit.

It is well known that the $\N=1^*$ and $\N=2^*$ theories are related to the
elliptic Calogero-Moser system; in particular, the Coulomb branch of
the $\N=2^*$ theory is the spectral curve of the integrable system and
the elliptic superpotential that arises on breaking to $\N=1^*$ 
is one of the Hamiltonians. In this paper,
we ask whether the matrix model results of \cite{cft} can be used to
deduce whether this relation to integrable systems is maintained under
the Leigh-Strassler $q$ deformation? The answer is yes, since it
transpires that the
Leigh-Strassler $q$ deformation involves a natural one parameter
deformation of the elliptic Calogero-Moser system known as the
Ruijsenaars-Schneider system. This result allows us to solve the
Seiberg-Witten theory of the 
Leigh-Strassler $q$ deformation of the $\N=2^*$ theory. In other words,
we find the exact description of this $\N=1$ Coulomb branch.

\section{Review of the relevant deformations of $\N=4$}

In this section we will review the story of the 
relevant deformations of $\N=4$. In other words, we take $\lambda=1$
and $\beta=0$ in \eqref{tree}. This theory is known as the $\N=1^*$ theory.
The vacuum structure of this theory was originally determined by 
compactifying the theory on ${\mathbb R}^3\times S^1$ \cite{nick} (see
also \cite{us}). In
order to describe how this works it is useful to think of the $\N=1^*$
deformation in two stages: firstly with only the mass $m$ non-zero,
which describes the so-called $\N=2^*$ theory, and then with the final
mass $\mu$ turned on.

The $\N=2^*$ theory in four dimensions has a Coulomb branch which was
described by Donagi and Witten \cite{donwitt}. In particular, it is the
moduli space of the Seiberg-Witten curve which in this case is a
certain $N$-fold ramified cover of the basic torus $E_\tau$ with complex
structure $\tau$ the complex gauge coupling. In the following, we
choose a normalization in which the periods of $E_\tau$ are:
\EQ{
2\omega_1=2\pi i\ ,\qquad2\omega_2=2\pi i\tau\ . 
}
The Seiberg-Witten curve is also the
spectral curve of the Calogero-Moser integrable system
\cite{Martinec:1995by,Martinec:1995qn,Martinec:1995ne} describing the
interaction $N$ particles according to the Hamiltonian
\cite{Calogero:1975ii,Moser} 
\EQ{
H=\sum_{a}p_a^2+m^2\sum_{a\neq b}\wp(x_a-x_b)\ ,
\label{ham}
}
with momenta $p_a$ and positions $x_a$. In the following, we shall
freeze the trivial centre-of-mass motion by choosing 
\EQ{
\sum_ap_a=\sum_ax_a=0\ .
}
This amounts to restricting to the $\SU(N)$ gauge group without an
additional $\U(1)$ factor. In the above, $\wp(x)$ is the
Weierstrass function defined on the torus $E_\tau$.

Integrability is manifest in the Lax formalism. One
defines the Lax matrix with elements
\EQ{
\phi_{ab}(z)=p_a\delta_{ab}+im(1-\delta_{ab})
\frac{\sigma(x_{ab}+z)}{\sigma(x_{ab})\sigma(z)}\ ,
\label{lax}
}
where $\sigma(z)$ is the Weierstrass sigma function and $x_{ab}=x_a-x_b$.
The time evolution can then be written via another matrix $M$:
\EQ{
\dot\phi=[M,\phi]
}
which clearly leaves invariant the spectrum of $\phi(z)$. In
particular, the spectral curve $\Sigma$
\EQ{
F(v,z)={\rm det}\big(v1_{\sst[N]\times[N]}-\phi(z)\big)=0
\label{spcurve}
}
is left invariant. This curve is an $N$-fold branched
covering of the basic torus $E_\tau$:
\EQ{
F(v,z+2\omega_1)=F(v,z+2\omega_2)=F(v,z)
}
and plays the r\^ole of the Seiberg-Witten curve for the theory.
A basis for the space of $N-1$ Hamiltonians is obtained by taking the finite
parts of ${\rm Tr}\,\phi^k(z)$, $k=2,\ldots,N$, around the poles at $z=0$.
In particular,
\EQ{
{\rm Tr}\,\phi^2(z)=-m^2N(N-1)\wp(z)+H\ ,
}
where $H$ is \eqref{ham}.

The Coulomb branch of the $\N=2^*$ theory is identified with the 
moduli space of the spectral curve \eqref{spcurve} of the
complexified Calogero-Moser system---so $x_a$ and $p_a$ are taken to be
complex---parameterized by the $N-1$, now complex, Hamiltonians. The
relation with the integrable system becomes even more satisfying once the
theory is compactified on a circle: in that case the Coulomb branch
doubles in dimension because there are $N-1$ Wilson lines and dual
photons which can be amassed into $N-1$ additional complex scalar
fields. These variables are naturally valued on a multi-dimensional
torus which is nothing but the space of the angle variables of the
Calogero-Moser system conjugate to the Hamiltonian, or action,
variables. The multi-dimensional torus is naturally identified with
the Jacobian variety ${\EuScript J}(\Sigma)$ of the spectral curve.
So in the compactified theory the Coulomb branch is naturally
identified with the whole phase space of the integrable system. 

Now we
can turn on the final mass deformation $\mu$. This lifts the Coulomb
branch according to a superpotential which is precisely the quadratic
Hamiltonian $H$ in \eqref{ham}. This is the exact elliptic
superpotential of \cite{nick}. Actually more general deformations of
the form $\Tr\,W(\Phi)$ can be considered as these simply correspond to
some linear combination of the $N-1$ Hamiltonians. 

One way to understand why the superpotential
is exact and to more fully elucidate the relation with the
integrable system is to realize the whole set-up within string
theory. We briefly describe the chain of arguments.
One starts with Witten's elliptic brane construction in Type IIA
string theory \cite{wittm}. The background spacetime is
${\mathbb R}^9\times S^1$. There are $N$ D4-branes whose world-volume lies
in ${\mathbb R}^4\times S^1$ parameterized by $x^n$, $n=0,1,2,3,6$,
where $x^6$ is the periodic coordinate. 
There is one NS5-brane with a world-volume along $x^n$,
$n=0,1,2,3,4,5$. The low energy theory on the D4 branes is then
four-dimensional $\N=4$ supersymmetric gauge theory. Up until now, the
single NS5 plays no r\^ole. However by twisting the spacetime we can
break to $\N=2^*$. This is achieved by a non-trivial fibration of  the complex direction 
$v=x^4+ix^5$ over the $x^6$ circle:
\EQ{
x^6\longrightarrow x^6+2\pi L\ ,\qquad v\longrightarrow v+m\ .
}
In this case the D4-branes have to split at the NS5-brane and the
resulting theory at low energies is the $\N=2^*$ theory. In order to
include quantum corrections one now lifts the configuration to M
theory \cite{wittm}. A new dimension appears parameterized by $x^{10}$
which, along with $x^6$, forms a torus $E_\tau$ with complex structure $\tau$,
the underlying complex gauge coupling of the theory. In M theory the
configuration of D4-branes and NS5-brane lifts to a single M theory
5-brane with a world volume ${\mathbb R}^4\times\Sigma$, where 
$\Sigma$ is a 2-surface embedded non-trivially in the four-dimensional
space ${\mathbb R}^2\times E_\tau$, parameterized by the two complex
coordinates $(v,z)$. The embedding is described by the Seiberg-Witten
curve and has the form
\EQ{
F(v,z)=v^N-f_1(z)v^{N-1}+f_2(z) 
v^{N-2}-\cdots+(-1)^Nf_{N}(z)=0\ .
\label{swc}
}
The functions $f_a(z)$ are elliptic on the torus $E_\tau$ (which we
take, as before, to have periods $(2\omega_1=2\pi i,2\omega_2=2\pi
i\tau)$), with the following analytic structure. The elliptic
functions $f_a(z)$ have a pole of order $a$ at $z=0$ and upon a
suitable shift in $v$, the singularities of $F(v,z)$ can be converted
into a simple pole at $z=0$. It is not difficult to show
\cite{donwitt,wittm} that 
the curve \eqref{swc} is precisely the spectral curve of the
Calogero-Moser system \eqref{spcurve}. Note that in order to describe
$\SU(N)$, rather than $\U(N)$, we decouple the 
centre-of-mass motion and this sets $f_1(z)=0$.

As an alternative to lifting to M theory, following Kapustin
\cite{kap}, we take a different route that leads to the same
result but is more suitable for our needs. The idea is to compactify
one of the spacetime directions of the D4-branes, say $x^3$, on a
circle of radius $R$. For small radius $R$,
we can now perform a T-duality in $x^3$ to yield the Type IIB
configuration of D3-branes spanning $x^0,x^1,x^3,x^6$. Under this
duality, the string coupling is transformed to $g'_s=\sqrt{\alpha'}/R$.
We follow this
with an $S$-duality on the four-dimensional theory on the
D3-branes. Finally, we perform, once again, a T-duality in $x^3$ to return to a
Type IIA configuration with D4-branes spanning
$x^0,x^1,x^2,x^3,x^6$. However, due to the intervening $S$-duality, the
radius of the $x^3$ is not returned to its original value. The new
radius is $Rg'_s=g_s\sqrt{\alpha'}$. In other words, it is independent
of the radius $R$.\footnote{All memory of $R$ is not lost because
the string coupling in the dual theory is $R^{2}/(g_s\alpha')$.}
The theory describing the collective dynamics
of these D4-branes is the mirror dual, or ``magnetic'', theory. It is a
five-dimensional theory compactified on ${\mathbb R}^3\times T^2$.
The most significant fact is that the torus
$T^2$ has complex structure $\tau$, the complex gauge coupling of the
original theory, and so is identified with the basic torus $E_\tau$
but now realized in the $(x^3,x^6)$ space rather than the
$(x^6,x^{10})$ space of the M theory construction. 

The discussion so far has been simplified because we have ignored the
fact that there is an NS5-brane in the original Type IIA set-up on
which the D4-branes can split when the mass $m$ is non-vanishing. 
Under the first T-duality the NS5-brane become a Type IIB NS5-brane. Then
under $S$-duality it becomes a D5'-brane (to distinguish it from
the other D-branes in the problem). Finally the T-duality around
$x^3$ changes it into an D4'-brane spanning
$x^0,x^1,x^2,x^4,x^5$, but localized at points on the
$(x^3,x^6)$ torus. As usual in a mirror transform we have mapped
the Coulomb branch of the original theory,
where the D4-branes were prevented from moving off the NS5-branes, to
the Higgs branch of the magnetic theory. 

The configuration that we are considering preserves eight real
supersymmetries. So we have a realization of the Coulomb branch of the
3-dimensional theory as the Higgs
branch of an ``impurity'' gauge theory with eight real
supercharges. This is why the mirror map is a useful device. The Higgs branch
will not be subject to quantum corrections and in this way we are able to
``solve'' the theory. It is naturally described by a set of
$D$- and $F$-flatness equations which involve the, suitably normalized,
components of the dual $\SU(N)$
gauge field $\tilde A_{z,\bar z}=\tfrac12(\tilde A_3\pm i\tilde A_6)$
of the D4-branes along the torus\footnote{We 
perform an overall re-scaling of the torus $T^2$ so that it
becomes precisely $E_\tau$ parameterized by the holomorphic
coordinate $z$ with periods $2\omega_1=2\pi i$ and 
$2\omega_2=2\pi i\tau$.} and the adjoint-valued
complex scalar field $\phi$ describing the fluctuations of
the D4-branes in the $x^4,x^5$ direction. 
In addition, the D4'-brane impurity
gives rise to a hypermultiplet $(Q_a\,\tilde Q_a)$ transforming in the
$(\BN,\overline{\BN})$-representation of $\SU(N)$, which is {\it localized\/}
at a point on the torus which we choose at $z=0$. The $D$- and $F$-flatness
conditions, respectively, with some convenient
choice of normalization of the hypermultiplets, read \cite{us,kap}
\AL{
&\Big(\tilde F_{z\bar
z}-[\phi,\phi^\dagger]\Big)_{ab}=-i\pi\delta^2(z,\bar z)
(Q_aQ^\dagger_b-\tilde Q^\dagger_a\tilde Q_b)\ ,\label{mma}\\
&\Big(\tilde{\cal D}_{\bar z}\phi\Big)_{ab}=
i\pi\delta^2(z,\bar z)(Q_a\tilde Q_b-m\delta_{ab})\ .
\label{mmb}
}
Here, \eqref{mma} is a real equation and \eqref{mmb} is a complex
equation and $\tilde {\cal D}_{\bar z}\phi=\partial_{\bar
z}\phi+[\tilde A_{\bar
z},\phi]$.  Notice how the mass enters into the $F$-flatness
equation. These equations are a generalization of Hitchin's
self-duality equations reduced to two dimensions
\cite{hitchin}. 

The Coulomb branch of the 3-dimensional theory 
is then the solution of the $D$- and $F$-flatness  
conditions modulo local $\SU(N)$ gauge transformations on the
torus. The construction is an example of an infinite
hyper-K\"ahler quotient and so the Coulomb branch of the 3-dimensional
theory is a
$4(N-1)$-dimensional hyper-K\"ahler space. 
As usual as long as we are interested in holomorphic quantities we can
relax the $D$-flatness condition and then solve for the $F$-flatness
condition moduli complex local gauge transformations---those valued in 
$\SL(N,{\mathbb C})$. 

To proceed, it is very convenient to use up (most of) the local part
of the quotient group, $\SL(N,{\mathbb C})$, to transform the
anti-holomorphic component $A_{\bar
z}$ into a constant diagonal matrix:
\EQ{
\tilde A_{\bar z}=\frac{\pi i}{2(\bar\omega_2\omega_1-\bar\omega_1\omega_2)}
{\rm diag}(x_1,\ldots,x_N)
\label{gauge}
}
with $\sum_{a=1}^Nx_a=0$.
The only local transformations that remain act by shifting the
$x_a$ by periods of the torus:
\EQ{
x_a\to x_a+2n\omega_1+2m\omega_2\ ,\qquad m,n\in{\mathbb Z}\ .
}
The remaining global part of the gauge group is also fixed, up to
permutations of the $x_a$, by choosing
\EQ{
\tilde Q_a=1\ .
\label{gauge2}
}
We can now solve explicitly for $\phi$ to get a very concrete parameterization
of the Coulomb branch. With $\tilde A_{\bar z}$ diagonal, 
the diagonal elements $\phi_{aa}$ are meromorphic functions on the
torus with a possible simple pole at $z=0$. However, there are no such
functions other than a constant; consequently,
\EQ{
\phi_{aa}=p_a\qquad\text{and}\qquad Q_a=m\ .
\label{ggf}
}
where the $p_a$ with $\sum_{a=1}^Np_a=0$ are new parameters.  

The off-diagonal elements are
\EQ{
\phi_{ab}(z,\bar z)=im
\frac{\sigma(x_{ab}+z)}{\sigma(x_{ab})\sigma(z)}e^{\psi(z,\bar z)x_{ab}}
\qquad(a\neq b)\ .
\label{ggg}
}
Here $x_{ab}\equiv x_a-x_b$, and we have
defined
\EQ{
\psi(z,\bar z)=\frac1{\bar \omega_2\omega_1-\bar
\omega_1\omega_2}\big[\zeta(\omega_2)
(\bar \omega_1 z-\omega_1\bar z)-\zeta(\omega_1)(\bar
\omega_2z-\omega_2\bar z)\big]\ .
}
One can readily verify that $\phi_{ab}(z,\bar z)$ is periodic on
the torus. Furthermore, a shift of $x_a$ by a lattice vector
$2\omega_\ell$, can be undone by a large gauge transformation on the
torus as anticipated earlier. Up to a simple diagonal 
gauge transformation,
\EQ{
U_{ab}=e^{-\psi(z,\bar z)x_a}\delta_{ab}\ ,
\label{gat}
}
the matrix $\phi$, with elements \eqref{ggf} and \eqref{ggg},
is equal to the Lax matrix of the elliptic
Calogero-Moser system \eqref{lax} where the $x_a$ are the positions
and the $p_a$ are the momenta.

We now have an explicit
parameterization of the 3-dimensional Coulomb branch 
furnished by $\{p_a,x_a\}$ with
\EQ{
\sum_ap_a=\sum_ax_a=0\ .
\label{rouo}
}
As we have already alluded to above,
there is also a completely integrable dynamical
system for which $x_a$ are the positions and $p_a$ are momenta with
the usual Poisson bracket structure.
It is the elliptic Calogero-Moser system \cite{Calogero:1975ii,Moser}. 
In particular, as we have already stated, the spectral
curve \eqref{spcurve} is
precisely the Seiberg-Witten curve $\Sigma$ of the
four-dimensional theory before compactification to three dimensions.
Since the dynamical system is completely integrable, there are $N-1$ (complex)
Hamiltonians. These are identified with coordinates on the
Coulomb branch of the four-dimensional theory.
The conjugate angle variables---also complex---take 
values in the Jacobian of $\Sigma$. 

A basis in the space of the Poisson-commuting 
Hamiltonians of the dynamical system is obtained by taking
the gauge invariant quantities ${\rm Tr}\phi^k(z)$, $k=2,\ldots,N$,
for some fixed $z$. These Hamiltonians parameterize the Coulomb branch
of the four-dimensional theory. In particular, the quadratic
Hamiltonian is the finite part of 
\EQ{
{\rm Tr}\,\phi(z)^2=-N(N-1)m^2\wp(z)+\sum_{a=1}^Np_a^2+m^2\sum_{a\neq
b}\wp(x_{ab}) 
}
around $z=0$.

Now that we have established the description of the Coulomb branch of
the compactified theory in terms of the Calogero-Moser integrable
system, we can break to $\N=1^*$ by adding the general
perturbation
\EQ{
\sum_{k=2}^N\mu_k{\rm Tr}\,\Phi^k 
}
to the tree-level superpotential. In the Higgs branch description of
the compactified theory this gives rise to an exact ``elliptic''
superpotential which is some linear combination of Hamiltonians:
\EQ{
W_{\rm eff}(x_a,p_a)=\sum_{k=2}^N\mu_k H_k\ .
}
This superpotential lifts the three-dimensional Coulomb branch and
importantly is independent of the compactification radius and so
is equally valid in the four-dimensional limit. The only subtlety
involved is in identifying the coordinates $\{H_k\}$ in the space of
Hamiltonians. This problem involves resolving operator mixing
ambiguities \cite{nick,us,Aharony:2000nt}. 
For the quadratic perturbation $\mu{\rm Tr}\,\Phi^2$ 
the situation is simple and the exact elliptic superpotential is
\EQ{
\frac1{\mu}W_{\rm eff}(x_a,p_a)=\sum_{a=1}^Np_a^2+m^2\sum_{a\neq
b}\wp(x_{ab}) 
}
up to an additive constant which is not physically significant.

We can go on to consider the vacuum structure of the theory by
extremizing $W_{\rm eff}(x_a,p_a)$. First of all, it is clear that in any
vacuum $p_a=0$. There are two classes of vacua: the massive and
massless. The former have been completely determined \cite{nick} 
while the classification of the 
latter is still an unsolved problem \cite{Aharony:2000nt}. The massive
vacua have a very beautiful interpretation from the point-of-view of
the dynamical system \cite{mm1,us}: they are precisely {\it equilibrium
configurations\/} with respect to the space of flows defined by
the $N-1$ Hamiltonians.\footnote{Here, ``time'' is an
auxiliary concept referring to
evolution in the dynamical system and not a spacetime concept in the
field theories under consideration.}
The point is that the massive vacua correspond to points of the
four-dimensional Coulomb branch for which $\Sigma$ degenerates to a
torus: cycles pinch off and one is
left with an $N$-fold un-branched cover of the basic torus 
$E_\tau$. This means that the
Jacobian Variety ${\EuScript J}(\Sigma)$ itself
degenerates: at these points the period matrix only has rank 1, with
non-zero eigenvalue $\tau$. The remaining torus is associated with the
centre-of-mass motion of the integrable system, so the overall $\U(1)$
factor, which we have removed by
\eqref{rouo}.
So at a massive vacuum, the remaining angle variables must stay fixed
under any time evolution. Since the Hamiltonians are by
definition constants of the motion, this means that the entire
dynamical system must be static at a massive vacuum and the system is
at an equilibrium point. Consequently, a
massive vacuum is not only a critical point of the quadratic Hamiltonian
but simultaneously of all the other $N-2$ Hamiltonians. 

The simplest kind of massive vacua are labelled by two integers $p$ and $q$
with $pq=N$. All the other cases can be generated from these by 
modular transformations of $\tau$ (in fact all the massive vacua lie
on a single orbit of the modular group). The critical point is then \cite{nick}
\EQ{
x_a\in\Big\{
\frac{2r}{q}\omega_1+\frac{2s}{p}\omega_2,\quad 0\leq r<q,\ 0\leq s<p\Big\}
.
\label{bcase}
}
The proof that this is a critical point of $W_{\rm eff}$ is delightfully
simple. One only needs to use the fact that $\wp'(z)$ is an
odd elliptic function. Terms in the
sum $\sum_{b(\neq a)}\wp'(x_{ab})$ either cancel in pairs or vanish
because $x_{ab}$ is a half-lattice point. As we mentioned, the set
\eqref{bcase} does not exhaust the set of massive vacua.
For a given pair
$(q,p)$ we can generate $q-1$ additional vacua by replacing
$\tau\to\tau+l/p$, $l=0,\ldots,q-1$.
So the total number of massive vacua is equal to $\sum_{p|N}p$, as
expected on the basis of a semi-classical analysis \cite{nick,donwitt}.

In the vacua \eqref{bcase} 
one can write down an expression for the superpotential in terms
of the $2^{\rm nd}$ Eisenstein series:
\EQ{
W_{\rm eff}=-\frac{\mu m^2Np^2}{12}E_2(p^2\tau/N)\ ,
\label{weff}
}
up to a vacuum-independent constant.

\section{The Leigh-Strassler deformation}

Now we consider the deformed theory. We start
with the expressions that generalize the superpotential in a subset
of the  massive vacua \eqref{weff} that we derived from the matrix
model formalism \cite{cft}:
\EQ{
W_{\rm eff}={pN\mu M^2\over2\lambda^2\sin\beta}\cdot
{\theta_1^\prime(p\beta/2|p^2\hat\tau/N)\over \theta_1({p\beta}/2
\vert{p^2\hat\tau/N})}-{N\mu M^2\over4\lambda^2\sin^2\frac\beta2}\ .
\label{exv}
} 
In this expression, we have defined the renormalized gauge coupling
\EQ{
\hat\tau= \tau-{iN\over\pi}\ln\lambda\ .
\label{tren}
}
In the limit $\beta\to0$ and $\lambda\to1$ \eqref{exv} reduces to \eqref{weff}.

Using \eqref{exv} we will attempt to reverse the route followed in the
last section. First of all, in \cite{cft} we identified the following
elliptic superpotential for which the configurations \eqref{bcase} are
still critical points and for which the superpotential takes the
values \eqref{exv} up to an additive vacuum-independent constant. The
relevant expression is
\EQ{W_{\rm eff}(x_a)={im^2\mu\over 2\lambda^2\sin\beta}
\sum_{a\neq b}
\left(\zeta(x_a-x_b+i\beta)-
\zeta(x_a-x_b-i\beta)
\right)\ ,
\label{esp}
}
where $\zeta(z)$ is the Weierstrass zeta-function which can be defined via
$\wp(z)=-\zeta^\prime(z)$ where $\wp(z)$ is the Weierstrass function
for the torus $E_{\hat\tau}$ with periods $2\omega_1=2\pi i$ 
and $2\omega_2=2\pi i\hat\tau$. (Definitions of the elliptic functions
that arise can be found in standard texts, for example \cite{WW}.)

The question is whether $W_{\rm eff}(x_a)$ can be identified with a
Hamiltonian of a known integrable system? The answer is yes, when one
investigates the known integrable systems one finds a natural
candidate which is a one parameter deformation of the Calogero-Moser
system. It is known as the Ruijsenaars-Schneider system 
\cite{Ruijsenaars:vq,Ruijsenaars:1986pp,Ruijsenaars:pv} or sometimes known as
the relativistic elliptic Calogero-Moser system. The
first and second Hamiltonians can be written
\EQ{
H_1=\sum_a\rho_a\ ,\qquad
H_2=\sum_{a\neq b}\rho_a\rho_b\frac1{\wp(i\beta)-\wp(x_{ab})}\ .
\label{rsh}
}
The $\rho_a$ are not directly the momenta conjugate to $x_a$, 
in fact these are $p_a$ defined via
\EQ{
e^{p_a}=\rho_a\prod_{b(\neq
a)}\frac1{\sqrt{\wp(x_{ab})-\wp(i\beta)}}\ .
\label{momd}
}
In order to relate this system to our superpotential \eqref{exv}, we
will impose the constraint 
\EQ{
\sum_{a}\rho_a=N
\Big(\frac{im^2\wp'(i\beta)}{2\lambda^2\sin\beta}\Big)^{1/2}\ ,
\label{conr}
}
which along with $\sum_ax_a=0$, is the analogue of freezing out the
centre-of-mass motion that we did in the $\N=1^*$ in order to describe
the $\SU(N)$, rather then $\U(N)$, theory. Notice 
that this constraint is {\it not\/} the
same as freezing out the centre-of-mass motion of the
Ruijsenaars-Schneider system since this would be
$\sum_ap_a=0$. However, we will shortly show how this constraint
arises naturally.

It is then straightforward to show that $H_2(x_a,\rho_b)$, subject to the
constraint \eqref{conr}, is equal to \eqref{exv}, 
up to an additive vacuum-independent constant, once the $\rho_a$ are
integrated out. This procedure gives
\EQ{
\rho_a=\Big(\frac{im^2\wp'(i\beta)}{2\lambda^2\sin\beta}\Big)^{1/2}\ .
}
In order to complete the equality with \eqref{exv} one uses the
elliptic function identity
\EQ{
\frac{\wp'(i\beta)}{\wp(i\beta)-\wp(z)}=\zeta(z+i\beta)-\zeta(z-i\beta)-
2\zeta(i\beta)\ .
}
By writing
\EQ{
\rho_a=\Big(\frac{i\wp'(i\beta)}{2\lambda^2\sin\beta}\Big)^{1/2}\big(
m+i\tilde p_a\sigma(i\beta)\big)\ ,
}
for new coordinates $\tilde p_a$, with $\sum_a\tilde p_a=0$, and then taking 
the limit $\beta\to0$ (and $\lambda\to1$), 
the $2^{\rm nd}$ Hamiltonian $H_2$
reduces to the quadratic Hamiltonian of the elliptic Calogero-Moser
with momenta $\tilde p_a$.
We remark that this is apparently different from the usual limit of the
Ruijsenaars-Schneider system that gives the elliptic Calogero-Moser
system. In that limit, one takes $p_a\sim{\cal O}(\beta)$ and 
it is the first Hamiltonian $H_1$ that gives the
quadratic Hamiltonian of the elliptic Calogero-Moser system. 

It is clear that these facts identify the Coulomb branch of the
3-dimemsional $\SU(N)$ theory, {\it i.e.\/}~the theory with tree-level
superpotential
\EQ{W_{\rm cl}=
\Tr\big(i\lambda\Phi[\Phi^+,\Phi^-]_\beta+m\Phi^+\Phi^-\big)\ ,
}
as the spectral curve of the 
Ruijsenaars-Schneider system. The latter can be constructed 
by a simple deformation of
the 
Hitchin system description of the elliptic Calogero-Moser system described
in the last section. The idea is to demand that the Lax matrix
$\phi(z,\bar z)$
is no longer periodic on the underlying torus $E_{\hat\tau}$ (with complex
structure $\hat\tau$) rather there is a non-trivial boundary
condition:
\EQ{
\phi(z+2\omega_1)=\phi(z)\ ,\qquad\phi(z+2\omega_2)=e^{i\beta}\phi(z)\
.
\label{twist}
}
We can solve the $F$-flatness condition \eqref{mmb} modulo complex gauge
transformation as before but now incorporating the boundary condition
\eqref{twist}. Choosing the gauge \eqref{gauge} and \eqref{gauge2}, as
before, the solution for the elements of the Lax matrix is
\EQ{
\phi_{ab}(z,\bar z)=i(Q_a-m\delta_{ab})
\frac{\sigma(x_{ab}-i\beta+z)}{\sigma(x_{ab}-i\beta)\sigma(z)}
e^{\psi(z,\bar z)x_{ab}+i\beta\zeta(\omega_1)z/\omega_1}\ ,
\label{lrs}
}
with the tracelessness constraint, since we work in $\SU(N)$ rather
than $\U(N)$,
\EQ{
\sum_{a=1}^NQ_a=Nm\ .
\label{trle}
}
Notice that in contrast to the $\N=1^*$ case, the diagonal elements
are not constant and $Q_a$ is not forced to be $m$: in fact the $Q_a$
will be related to the conjugate momenta.

The quadratic Hamiltonian is extracted from the gauge invariant quantity
\SP{
{\rm
Tr}\,\phi^2(z)&=-e^{2i\beta\zeta(\omega_1)z/\omega_1}\frac{\sigma(z-i\beta)^2}
{\sigma(z)^2
\sigma(i\beta)^2}\sum_{ab}(Q_a-m\delta_{ab})(Q_b-m\delta_{ab})\frac{
\wp(x_{ab})-\wp(z-i\beta)}{\wp(x_{ab})-\wp(i\beta)}\\
&=e^{2i\beta\zeta(\omega_1)z/\omega_1}\frac{\sigma(z-i\beta)^2}
{\sigma(z)^2
\sigma(i\beta)^2}(\wp(i\beta)-\wp(z-i\beta))\\
&\qquad\qquad\times\Big\{-\frac{N(N-1)m^{2}}{\wp(i\beta)-\wp(z-i\beta)}+
\sum_{a\neq b}Q_aQ_b\frac1{\wp(i\beta)-\wp(x_{ab})}\Big\}\ .
}
We now recognize the second term in braces as proportional to the
Hamiltonian $H_2$ in \eqref{rsh} with 
\EQ{
\rho_a=\Big(\frac{i\wp'(i\beta)}{2\lambda^2\sin\beta}\Big)^{1/2}Q_a
}
and the \eqref{trle} is the traceless condition \eqref{conr}.
It is not so obvious that the Lax matrix \eqref{lrs} is equivalent to
the more conventional form in \cite{Ruijsenaars:1986pp}. Firstly, we
perform the diagonal gauge transformation \eqref{gat}.
Then we notice that the part involving $m$ is a constant proportional to the
identity matrix and so we can shift this away without affecting the
spectral curve or dynamics. Next, we have to
multiply by the overall factor
\EQ{
i\frac{\sigma(i\beta)\sigma(z)}{\sigma(z-i\beta)}
e^{-i\beta\zeta(\omega_1)z/\omega_1} 
}
and, finally, shift the spectral parameter $z\to z+i\beta$. The
resulting Lax matrix is then the one quoted in
\cite{Ruijsenaars:1986pp}:
\EQ{
\phi_{ab}=Q_a\frac{\sigma(x_{ab}-z)}{\sigma(x_{ab}-i\beta)}\ .
}
We remark that our construction of the Lax matrix via a
simple modification of the Hitchin system of the elliptic
Calogero-Moser system would appear to be simpler
than the alternative constructions in \cite{Arutyunov:1996uw,Arutyunov:1996vy}.

So we have demonstrated that the elliptic superpotential that
described the Leigh-Strassler deformed $\N=1^*$ is the quadratic
Hamiltonian of the Ruijsenaars-Schneider system. Furthermore, it now
follows that the Coulomb branch of the four-dimensional
Leigh-Strassler deformed $\N=2^*$ theory with tree-level
superpotential
\EQ{W_{\rm cl}=
\Tr\big(i\lambda\Phi[\Phi^+,\Phi^-]_\beta+m\Phi^+\Phi^- \big)
}
is described by the spectral curve of the
Ruijsenaars-Schneider system, {\it i.e.\/}~as in \eqref{spcurve} but
with the deformed Lax matrix \eqref{lrs}. We can write the curve as in
\eqref{swc}:
\EQ{
F(v,z)=v^N-f_1(z)v^{N-1}+f_2(z) 
v^{N-2}-\cdots+(-1)^Nf_{N}(z)=0\ ,
}
but with modified conditions of the $f_a(z)$. These functions have the
same pole structure as before; however they are no longer elliptic
functions rather they incorporate the non-trivial boundary conditions
on $E_{\hat\tau}$:
\EQ{
f_a(z+2\omega_1)=f_a(z)\ ,\qquad f_a(z+2\omega_2)=e^{ia\beta}f_a(z)\ ,
} 
so that 
\EQ{
F(v,z+2\omega_1)=F(v,z)\ ,\qquad F(v,z+2\omega_2)=F(ve^{i\beta},z)\ .
}
As before in the $\SU(N)$ theory $f_1(z)=0$. As an example, in the
$\SU(2)$ theory with a suitable re-scaling of $v$,
\EQ{
v\longrightarrow e^{2i\beta\zeta(\omega_1)z/\omega_1}\frac{\sigma(z-i\beta)^2}
{\sigma(z)^2
\sigma(i\beta)^2}v\ ,
} 
the spectral curve is
\EQ{
v^2+m^2+Q(2m-Q)\frac{\wp(i\beta)-\wp(z-i\beta)}{\wp(x)-\wp(i\beta)}=0\ ,
\label{spsut}
}
where $x\equiv x_{12}$ and $Q=Q_1=2m-Q_2$.

From this result it is now possible to identify how one must modify
Witten's brane construction of the $\N=2^*$ theory \cite{wittm} in order to
incorporate the Leigh-Strassler $q$ deformation. As one goes around the
$x^6$ circle one needs to incorporate a non-trivial rotation $v\to
e^{i\beta}v$. However, it is not so obvious how one can decouple the
overall $\U(1)$ factor in the brane set-up.

As a final comment, it is intriguing that the Ruijsenaars-Schneider
system also plays a r\^ole in the $\N=2^*$ theory lifted to five
dimensions and then compactified 
on a circle of radius $R$ \cite{Nekrasov:1996cz}. In this
case, the deformation parameter $\beta$ is identified with
$Rm$. However, apparently 
the resulting spectral curve is different because in
the five-dimensional case we have the constraint $\sum_ap_a=0$, where
the momenta are defined in \eqref{momd}, rather than the constraint
\eqref{conr} in the Leigh-Strassler case. 
For example, in the case of $\SU(2)$ the resulting spectral curve can
be written as 
\EQ{
v^2+2i\cosh p\sqrt{\wp(x)-\wp(iRm)}v+\wp(iRm)-\wp(z)=0\ ,
}
to compare with \eqref{spsut}.

\vspace{1cm}

I would like to thank Brett Taylor for suggesting some improvements
to the original manuscript.

\end{document}